\documentclass[]{aastex61}

\usepackage[normalem]{ulem}

\received{}
\revised{}
\accepted{}
\submitjournal{ApJ}

\shorttitle{High-resolution spectroscopy of V960 Mon}
\shortauthors{Takagi et al.}
\begin{document}

\title{Revealing the spectroscopic variations of FU Orionis object V960 Mon with high-resolution spectroscopy}

\correspondingauthor{Yuhei Takagi}
\email{takagi@naoj.org}

\author{Yuhei Takagi}
\affil{Subaru Telescope, National Astronomical Observatory of Japan, 650 North A`ohoku Place, Hilo, HI 96720, USA}

\author{Satoshi Honda}
\affil{Nishi-Harima Astronomical Observatory, Center for Astronomy, University of Hyogo, 407-2, Nishigaichi, Sayo, Sayo, Hyogo 679-5313, Japan}

\author{Akira Arai}
\affil{Koyama Astronomical Observatory, Kyoto Sangyo University, Motoyama, Kamigamo, Kita-ku, Kyoto City, Kyoto 603-8555, Japan}

\author{Jun Takahashi}
\affil{Nishi-Harima Astronomical Observatory, Center for Astronomy, University of Hyogo, 407-2, Nishigaichi, Sayo, Sayo, Hyogo 679-5313, Japan}

%\author{Kumiko Morihana}
%\affil{Division of Particle and Astrophysical Science, Graduate School of Science, Nagoya University, Furo-cho, Chikusa-ku, Nagoya, Aichi, 464-8602, Japan}

\author{Yumiko Oasa}
\affil{Faculty of Education, Saitama University, 255 Shimo-Okubo, Sakura, Saitama, Saitama 388-8570, Japan}

\author{Yoichi Itoh}
\affil{Nishi-Harima Astronomical Observatory, Center for Astronomy, University of Hyogo, 407-2, Nishigaichi, Sayo, Sayo, Hyogo 679-5313, Japan}



\begin{abstract}

We present the results of the high-resolution spectroscopy of the FU Orionis-type star V960 Mon.
The brightness of V960 Mon decreased continuously after the outburst was detected in 2014 November.
During this dimming event, we have carried out medium-resolution spectroscopic monitoring observations and found that the equivalent width of the absorption features showed variations.
To further investigate the spectroscopic variations, we conducted a high-resolution spectroscopic observation of V960 Mon with the Subaru Telescope and the High Dispersion Spectrograph
on 2018 January 8 and 2020 February 1. By comparing this spectrum with the archival data of the Keck Observatory and the High Resolution
Echelle Spectrometer taken between 2014 and 2017, we found that the absorption profiles changed as V960 Mon faded.
The line profile of absorption lines such as Fe~{\footnotesize I} and Ca~{\footnotesize I} can be explained by a sum of the spectra of the
disk atmosphere and the central star. The model spectrum created to explain the variations of the line profiles suggests that
the effective temperature of the central star is $\sim$5500 K, which is comparable to that of the pre-outburst phase with a distance of 1.6 kpc with Gaia.
The spectrum also shows that the effective temperature of the disk atmosphere decreased as V960 Mon faded.
The variations of the H$\alpha$ and Ca~{\footnotesize II} lines (8498.0 {\AA}, 8542.1 {\AA}) also show that the V960 Mon spectrum became central-star dominant.

\end{abstract}

\keywords{protoplanetary disks --- stars: formation --- stars: pre-main sequence}



\section{Introduction} \label{sec:intro}

An FU Orionis type-star (FUor) is a pre-main sequence star that shows a sudden increase of brightness in optical wavelength.
Understanding the mechanism of the FUor outbursts is crucial for revealing the star and planet formation processes.
The optical spectra of FUors show similar features to those of F to G type supergiants, whereas the near-infrared spectra resemble those of K- to M-type supergiants \citep[e.g.,][]{Hartmann1996,Audard2014}.
The absorption profiles show a double-peak or a flat-bottom shape \citep[e.g.,][]{Miller2011}.
The model of \citet{Hartmann1985}, which suggested that the outburst is triggered by an increase in the mass accretion rate,
\citep[from $10^{-8}$ -- $10^{-7}$~$M_{\odot}$~yr$^{-1}$ to $10^{-6}$ -- $10^{-4}$~$M_{\odot}$~yr$^{-1}$, ][]{Audard2014} reproduces these spectroscopic features.
The disk wind also contributes to the formation of absorption features \citep{Eisner2011}.

V960 Mon (2MASS J06593158-0405277) is an FUor first identified in 2014 November \citep{Maehara2014}.
The maximum magnitude was $\sim$3 mag brighter in the optical wavelength compared to the quiescent phase.
V960 Mon gradually faded by 0.5--1 mag between October 2014 and April 2015 \citep{Hackstein2015}, and it kept fading with a lower rate.
Spectroscopic observations of V960 Mon in the early phase of the outburst demonstrated typical characteristics of FUors.
The high-resolution optical spectrum \citep{Hillenbrand2014} showed blueshifted absorption components in Na D doublet, H$\alpha$, and Ca~{\footnotesize II} triplet lines, which
is evidence of active outflow. Other absorption lines were similar to those of F-type giant stars.

We conducted medium-resolution optical spectroscopic monitoring of V960 Mon to investigate the evolution of FUor outbursts \citep{Takagi2018}.
The Medium And Low-resolution Longslit Spectrograph \citep[MALLS;][]{Ozaki} equipped on the 2.0 m Nayuta Telescope at the Nishi-Harima Astronomical Observatory was used for the monitoring.
Spectra of V960 Mon were collected for 53 nights between with 2015 January 27 and 2017 January 31.
The wavelength coverage was 6280 -- 6750 {\AA} with the resolving power of $R\sim10000$.
During this observation period, we found variations in the strength of the absorption lines.
While the equivalent width of Fe {\footnotesize I} and Ca {\footnotesize I} lines were nearly constant, the peak depth of Fe {\footnotesize I} and Ca {\footnotesize I} lines became deeper.
Moreover, the equivalent width of the Fe {\footnotesize II} line (6456.4 {\AA}) decreased, and its peak became shallower.
The comparison of the equivalent width of these lines and the synthetic spectra suggest that the variations of the absorption lines
correspond to a decrease in effective temperature ($T_\mathrm{eff}$) and an increase in surface gravity ({\it g}).
%This result may suggest the evolution of the protoplanetary disk during the FUor outbursts.
These changes may indicate the variation of the protoplanetary disk during the FUor outburst, such as change of temperature distribution, mass accretion, and vertical structure.

To further study these spectroscopic variations of V960 Mon, we conducted high-resolution spectroscopy to investigate the changes in the line profile of the absorption lines.
In our previous mid-resolution spectroscopic study of V960 Mon, we assumed that the absorption lines have a single gaussian profile.
Because of the insufficient resolution power, it was difficult to investigate the variation of the line profiles.
The detailed evolution of V960 Mon outburst can be revealed by the line profile investigations.
Time series high-resolution spectra of FUors produce crucial information to understand the evolution of the wind and the disk during the outburst \citep[e.g.,][]{Lee2015}.
In section~2, we describe the details of the observed data and the archival data.
The brightness information of V960 Mon is summarized in section~3.
We discuss the change in both absorption lines and the emission lines and the cause of these variations in section~4.

\section{Observations and data reductions} \label{sec:obs}

\subsection{Subaru HDS}

The high-resolution spectrum of V960 Mon was obtained with the High-Dispersion Spectrograph \citep[HDS;][]{Noguchi2002} mounted on the optical Nasmyth focus of the Subaru Telescope.
Observations were conducted on 2018 January 8 (UT) and 2020 February 1 (UT) with averaged seeing sizes of $\sim1.5"$ and $\sim0.6"$, respectively.
To resolve the components of the absorption features, a spectrum with high resolution and high signal-to-noise ratio (S/N) was needed.
An image slicer with a slice pattern of $0".45\times3$ \citep{Tajitsu2012} was used to achieve high S/N with high spectral resolution ($R\sim80000$).
The exposure times were 6000 s (1500 s $\times$ 4) and 2400 s (1200 s $\times$ 2), respectively.
The nearby early-type star HR 2901 \citep[B9V; ][]{Houk1999} was also observed with an exposure time of 20 s before or after the observation of V960 Mon for the telluric line correction.
The angles of the echelle grating and the cross-disperser were set to observe the spectrum from 5850~{\AA} to 8550~{\AA} in order to obtain the absorption lines at $\sim$6000~{\AA}
and the Ca {\footnotesize II} line at 8542.1~{\AA} simultaneously.
Dispersed light was collected with two 2K $\times$ 4K CCDs with a pixel scale of 13.5$\mu$m.

The Image Reduction and Analysis Facility (IRAF) software package\footnote{IRAF is distributed by the National Optical Astronomy Observatory.} was used for reducing the data in the standard manner such as overscan subtraction,
%\sout{cosmic ray rejection,}
scattered light subtraction, flat fielding, and spectrum extract.
The comparison spectrum of Th-Ar was used for the wavelength calibration.
Before the telluric line corrections, the H$\alpha$ line and Ca {\footnotesize II} absorption features were removed from the HR 2901 spectrum.
To derive the line profiles of these broad lines, we corrected the blaze function of the aperture of which these lines are included,
by using the information of the blaze function of nearby echelle orders.
The telluric lines were then corrected with the spectrum of HR 2901 with no H$\alpha$ and Ca {\footnotesize II} lines.
The S/N of all V960 Mon spectra were calculated using the continuum regions (6505~--~6507.5 {\AA}, 6510~--~6512.5 {\AA}).
The observation details are shown in Table \ref{tab:log}.

\subsection{Keck HIRES Archive}

We used the archival data obtained with High Resolution Echelle Spectrometer \citep[HIRES;][]{Vogt1994} of Keck Observatory to investigate the time variations of the spectroscopic features (Table \ref{tab:log}).
Spectra were taken on seven nights between 2014 December and 2017 January (PI: L. Hillenbrand).
The wavelength range was from 4800 to 9200~{\AA}.
The slit width was set to $0".574$ or $0".861$, corresponding to the resolution power of 72000 and 48000, respectively.
Three 2K $\times$ 4K CCDs (blue, green, and red) with a pixel scale of 15~$\mu$m were used to collect the dispersed light.
We focused on the spectral features within 5850~{\AA} to 8550~{\AA} to discuss the spectral variations from 2014 December to 2020 February, which is the range overlapping with the Subaru HDS data.

We used the extracted data published on the website of the Keck Observatory Archive.
To investigate the uniformity of the data between the HDS data and HIRES data, and because the spectrum obtained with the green detector (6250~--~7750~{\AA}) was lacking in the public data of
2014 December 9 and 2016 February 2, we conducted the data reduction for data taken on these two nights.
The data of all three detectors were reduced in the same manner as HDS.
The spectra extracted from the blue and red CCDs with our reductions were comparable with the public data.
Therefore, we decided that the quality of the public data would be sufficient for our discussion.
The S/N of the spectra were calculated using the same continuum regions (6505~--~6507.5 {\AA}, 6510~--~6512.5 {\AA}) as the HDS data (Table \ref{tab:log}).

\renewcommand{\arraystretch}{0.95}
\begin{table}[t]
\caption{Summary of the high-resolution spectroscopic observations of V960 Mon.} \label{tab:log}
\begin{center}
\begin{tabular}{lcrrr}
  \hline \hline
  \multicolumn{1}{c}{Observation date}	& Instruments		& \multicolumn{1}{c}{Exp}	& \multicolumn{1}{c}{{\it R}}	& \multicolumn{1}{c}{S/N}	\\
  \multicolumn{1}{c}{(UT)}		& 	  		& \multicolumn{1}{c}{(s)}	& 				& 				\\ \hline
  2014 Dec 9--10  			& Keck HIRES        	& 915  				& 72000				& 131 \\
  2015 Feb 9        			& Keck HIRES        	& 731  				& 72000				& 130 \\
  2015 Oct 27       			& Keck HIRES        	& 300  				& 48000				& 89 \\
  2016 Feb 2        			& Keck HIRES        	& 600  				& 72000				& 142 \\
  2016 Oct 14       			& Keck HIRES        	& 180  				& 48000				& 80 \\
  2017 Jan 13       			& Keck HIRES        	& 180  				& 72000				& 67 \\
  2018 Jan 8        			& Subaru HDS        	& 6000 				& 80000				& 202 \\
  2020 Feb 1				& Subaru HDS		& 2400				& 80000				& 125 \\ \hline
 \hline
 \end{tabular}
 \end{center}
\end{table}
\renewcommand{\arraystretch}{1}

\section{Brightness variation} \label{sec:bri}

\begin{figure}[t]
 \begin{center}
   \includegraphics[width=8cm]{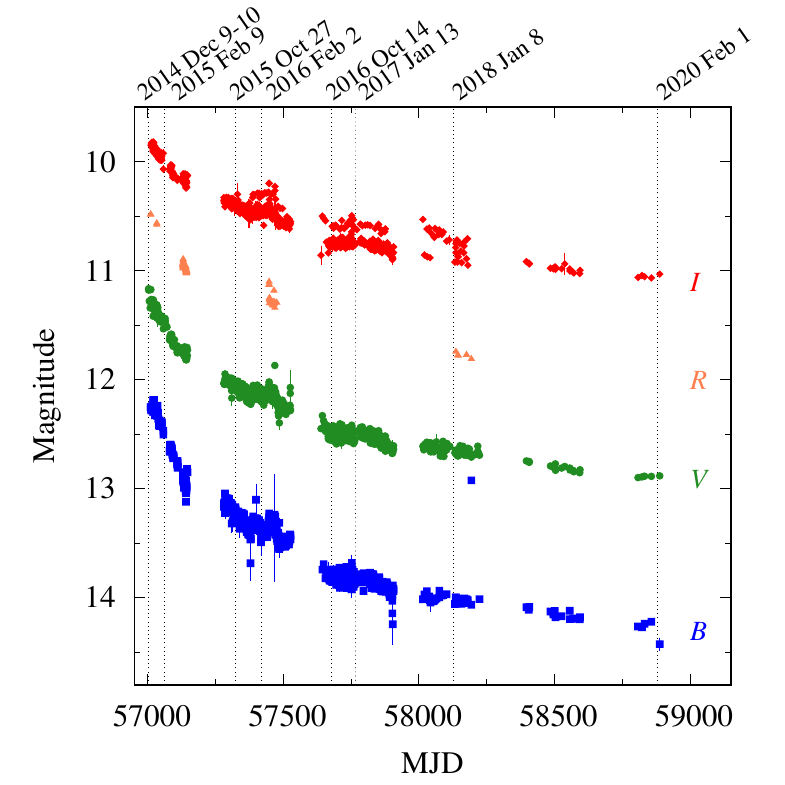}
 \end{center}
\caption{Light curve of V960 Mon with the data of AAVSO. The vertical dashed lines show the date when high-resolution spectra were obtained. \label{fig:lightcurve}}
\end{figure}

The information of brightness variation of V960 Mon during the period when the high-resolution spectra were observed is vital for understanding its spectroscopic variations.
The brightness of V960 Mon in the $r$- and $i$-bands in 2014 October were 2.9 mag brighter than those in the quiescent phase \citep{Hackstein2015}.
The light curve of V960 Mon during the period when the spectra were collected was created with data of the American Association of Variable Star Observers (AAVSO).
According to the light curve (Figure \ref{fig:lightcurve}), the magnitudes of $V$, $R$, and $I$-bands in the early phase of the outburst (2014 December 19) were $V=11.3$ mag, $R=10.5$ mag, and $I=9.8$ mag, respectively.
After the outburst, V960 Mon gradually faded.
The $V$- and $I$-band magnitudes of 2016 February 2 were 12.2 mag and 10.5 mag, respectively.
To estimate the $R$-band magnitude on this day, we used the information of the $V-R$ and $R-I$ colors.
Because these two colors were equal through the observation period, the $R$-band magnitude of 2016 February 2 was estimated as the average of $V$- and $I$-band magnitudes, which was 11.3 mag.
The $V$- and $I$-band magnitudes of 2018 January 8 were estimated the results of 2017 December 26 and 2018 January 22.
Interpolated $V$- and $I$-band magnitudes were 12.7 mag and 10.8 mag, respectively.
The $R$-band magnitude was estimated again as an average of $V$- and $I$-band magnitudes, which was 11.7 mag.
In the same manner, the brightness on 2020 February 1 was calculated as $V=12.9$ mag, $R=12.0$ mag, and $I=11.0$ mag.
Since the brightness of V960 Mon between October 2014 and December 2014 was nearly equal, the increment of $R$-band brightness can be estimated as 2.1 mag, 1.7 mag and 1.4 mag brighter on February 2016, January 2018, and February 2020, respectively, compared to the quiescent phase.

\section{Spectroscopic Variations} \label{sec:res}

\subsection{Line profiles of absorption features} \label{subsec:absprof}

\begin{figure*}[t!]
 \begin{center}
   \includegraphics[width=18cm]{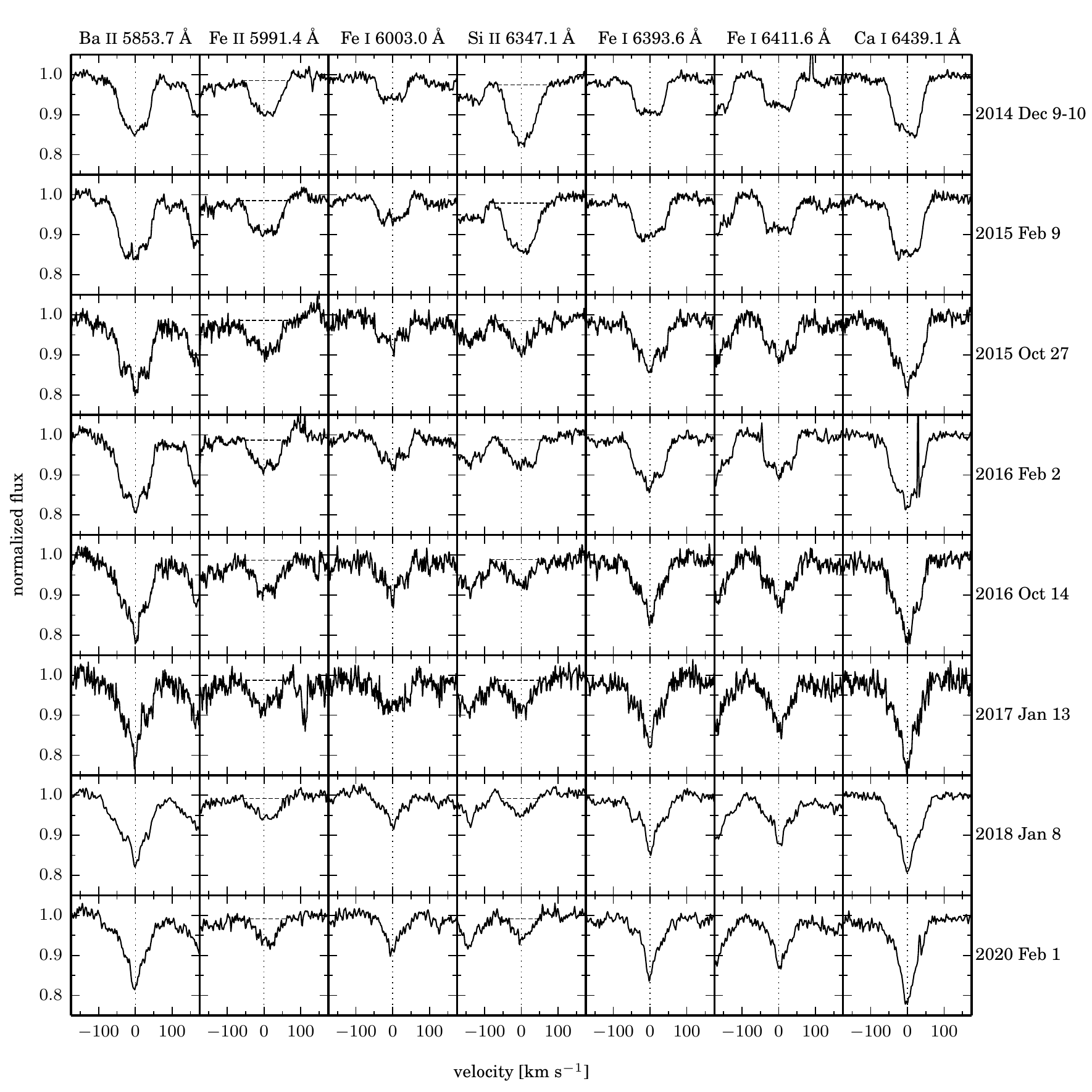}
 \end{center}
\caption{Line profiles of the typical absorption lines seen in V960 Mon spectra.
Horizontal dashed lines in the panels of Fe {\footnotesize II} (5991.4~{\AA}) and Si {\footnotesize II} (6347.1~{\AA}) lines show the
full width at 15 percent of the line center depth (see text for details). \label{fig:abscomp}}
\end{figure*}

In order to investigate the spectroscopic variation, we selected absorption lines within the observation
range of the Subaru/HDS data, which were not blended with adjacent absorption features (Table \ref{tab:linedata}).
The line variation patterns of these lines were not uniform (Figure \ref{fig:abscomp}).
Most of the lines such as Fe {\footnotesize I} and Ca {\footnotesize I} lines show a broad line profile with a flat bottom in the spectrum of 2014 December -- 2015 February.
This is one of the typical characteristics seen in FUors \citep{Petrov2008}.
As V960 Mon faded, Fe {\footnotesize I} and Ca {\footnotesize I} lines showed a peak around the line center (0~km~${\mathrm s}^{-1}$).
This narrow component became deeper as V960 Mon got fainter, whereas the broad component declined.

\renewcommand{\arraystretch}{0.95}
\begin{deluxetable}{lrrDc}
%\tablenum{4}
\tablecaption{Line data\label{tab:linedata}}
\tablewidth{0pt}
\tablehead{
\multicolumn1c{Line}	& \multicolumn1c{Wavelength}	& \multicolumn1c{L.E.P.}	& \multicolumn2c{log \it{gf}}	& \colhead{Ref.\tablenotemark{a}}	\\
\colhead{}			& \multicolumn1c{({\AA})}	& \multicolumn1c{(eV)}	& \multicolumn2c{}			& \colhead{}
}
\decimals
\startdata
Ba {\footnotesize II} & 5853.7 & 0.604 & $-$0.908 & 1 \\
Fe {\footnotesize I}	& 5934.7	& 3.929	& $-$1.12	& 1	\\
Fe {\footnotesize II}	& 5991.4	& 3.153	& $-$3.6	& 1	\\
Fe {\footnotesize I}	& 6003.0	& 3.882	& $-$1.120	& 2	\\
%Fe {\footnotesize I}	& 6065.5	& 2.609	& $-$1.530	& 1	\\
%Ni {\footnotesize I}	& 6108.1	& 1.676	& $-$2.44	& 1	\\
Ca {\footnotesize I}	& 6122.2	& 1.886	& $-$0.315	& 1	\\
Fe {\footnotesize I}	& 6265.1	& 2.176	& $-$2.550	& 1	\\
%Fe {\footnotesize I}	& 6344.1	& 2.433	& $-$2.923	& 1	\\
Si {\footnotesize II}	& 6347.1	& 8.121	& 0.149	& 1	\\
% Si {\footnotesize II}	& 6371.4	& 8.121	& $-$0.082	& 1	\\
Fe {\footnotesize I}	& 6393.6	& 2.433	& $-$1.576	& 1	\\
Fe {\footnotesize I}	& 6411.6	& 3.654	& $-$0.718	& 1	\\
% Fe {\footnotesize I}	& 6430.8	& 2.176	& $-$2.006	& 1	\\
% Fe {\footnotesize II}	& 6432.7	& 2.891	& $-$3.50	& 1	\\
Ca {\footnotesize I}	& 6439.1	& 2.526	& 0.47	& 1	\\
% Ca {\footnotesize I}	& 6449.8	& 2.521	& $-$0.55	& 1	\\
Ca {\footnotesize I}	& 6471.7	& 2.526	& $-$0.59	& 1	\\
Fe {\footnotesize I}	& 6546.2	& 2.759	& $-$1.536	& 1	\\
%Li {\footnotesize I}	& 6707.8	& 0.000	& $-$0.002	& 1	\\
%Li {\footnotesize I}	& 6707.9	& 0.000	& $-$0.303	& 1	\\
Ni {\footnotesize I}	& 7555.6	& 3.847	& $-$0.046	& 2	\\
Fe {\footnotesize I}	& 7568.9	& 4.283	& $-$0.882	& 2	\\
Ni {\footnotesize I}	& 7727.6	& 3.678	& $-$0.16	& 1	\\
Fe {\footnotesize I}	& 7937.1	& 4.312	& 0.152	& 2	\\ \hline \hline
\enddata
%\tablecomments{}
\tablenotetext{a}{References. (1) \citet{NIST}; (2) \citet{Kurucz1993}}
\end{deluxetable}
\renewcommand{\arraystretch}{1.0}

Absorption lines which are mainly seen in high $T_\mathrm{eff}$ stars such as Fe~{\footnotesize II} and Si~{\footnotesize II} did not show profile variations.
The depths of these lines became shallower as V960 Mon faded. In addition, no peak appeared at the line center.
The depth of Si {\footnotesize II} (6347.1 {\AA}) showed a significant decrease in the spectrum between February 2015 and October 2015,
while that of Fe~{\footnotesize II} (5991.4~{\AA}) decreased especially from 2017 January and 2018 January.
The depth variation of Fe~{\footnotesize I}, Ca~{\footnotesize I}, and Fe~{\footnotesize II} lines were comparable with the result of \citet{Takagi2018}.
The wavelength shift was not detected in the absorption features listed in Table \ref{tab:linedata}.

\begin{figure}[ht!]
 \begin{center}
   \includegraphics[width=8cm]{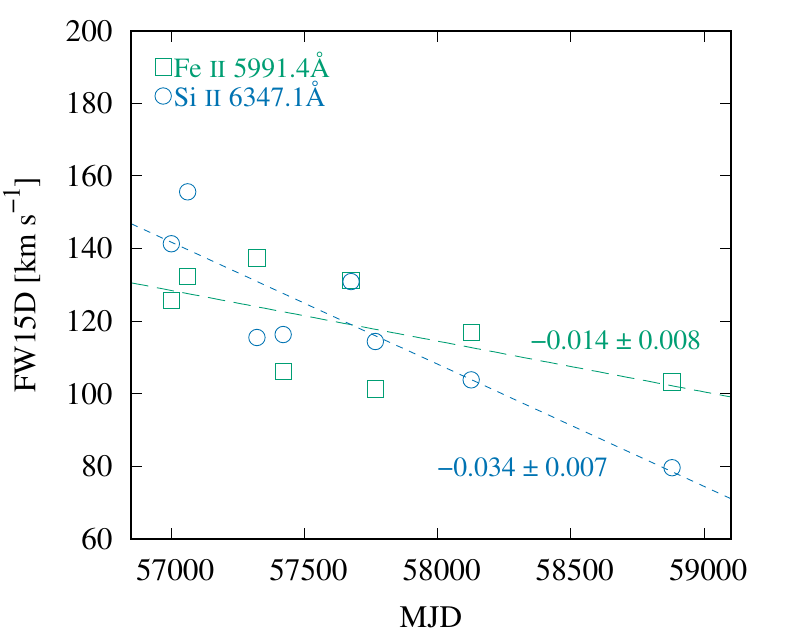}
 \end{center}
\caption{The line width variations of Fe {\footnotesize II} (5991.4~{\AA}) and Si {\footnotesize II} (6347.1~{\AA}) lines.
         The green long-dashed line shows the regression line of the Fe {\footnotesize II} (5991.4~{\AA}) with a slope of $-0.014\pm0.008$.
         The regression line of the Si {\footnotesize II} (6347.1~{\AA}) is represented with a blue short-dashed line with a slope of $-0.034\pm0.007$.
         \label{fig:fw15d}}
\end{figure}

To investigate the line width variation of the broad component, the full width of the line at the region close to the continuum level was measured \citep[e.g.,][]{Jayawardhana2003}.
In \citet{Jayawardhana2003}, the width at the 10 percent depth was employed.
But because the 10 percent widths of the observation data of V960 Mon were contaminated by the poor S/N and adjacent lines,
we measured the width of the lines at the 15 percent depth of the line center (FW15D; horizontal dashed lines in Figure \ref{fig:abscomp}).
Fe~{\footnotesize II} (5991.4~{\AA}) and Si~{\footnotesize II} (6347.1~{\AA}) lines were used to estimate the line width variations of the broad component (Figure \ref{fig:fw15d}).
The width of Si~{\footnotesize II} (6347.1~{\AA}) showed a decrease during the observation period, from $\sim$140~km~${\mathrm s}^{-1}$ to $\sim$80~km~${\mathrm s}^{-1}$.
The slope of the regression line calculated with the plots of Si~{\footnotesize II} (6347.1~{\AA}) was $-0.034\pm0.007$, which indicates the decreasing trend.
Meanwhile, that of Fe {\footnotesize II} (5991.4~{\AA}) was nearly constant or showed a slight decrease, of which the slope of the regression line was $0.014\pm0.008$.
This result indicates that the broad component, especially of Si {\footnotesize II}, became narrower as V960 Mon faded.

\subsection{Equivalent width of absorption lines} \label{subsec:absEW}

\begin{figure}[ht!]
 \begin{center}
   \includegraphics[width=8cm]{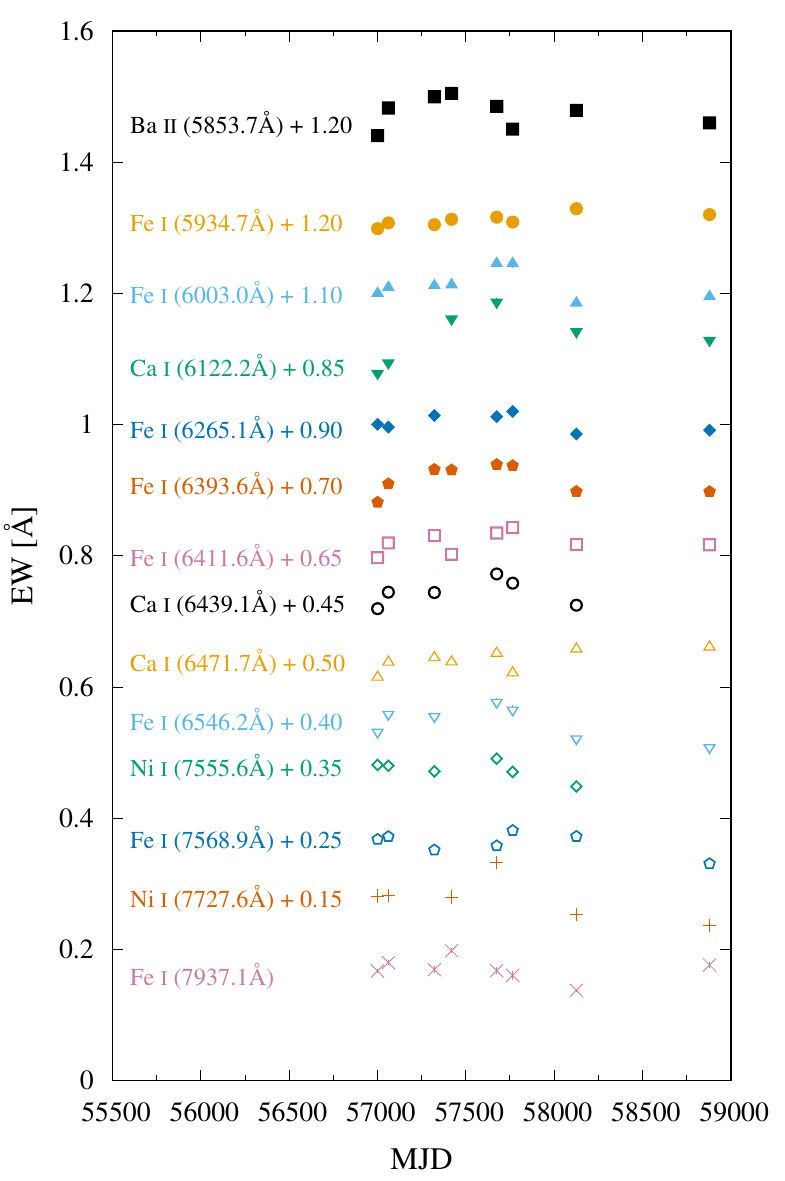}
 \end{center}
\caption{Equivalent widths of absorption lines. The offset EW values shown with the label for line name are added to each EWs.\label{fig:ew}}
\end{figure}

\begin{figure}[ht!]
 \begin{center}
   \includegraphics[width=8cm]{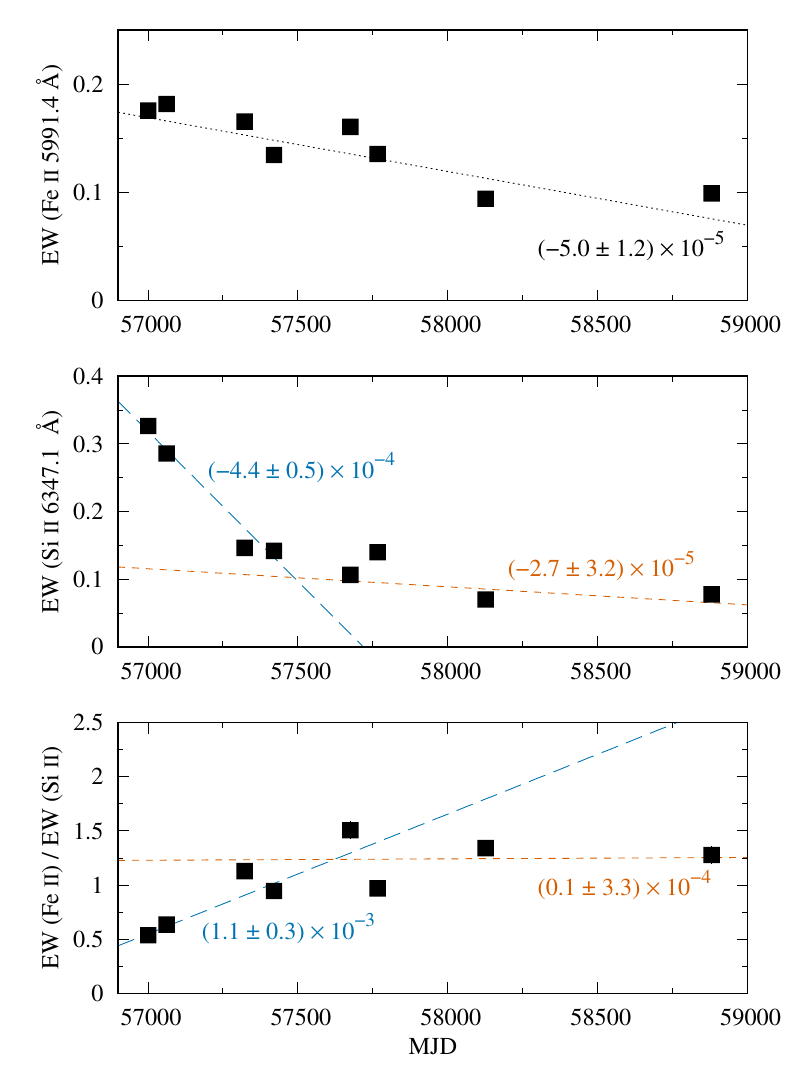}
 \end{center}
\caption{{\it Top}: Equivalent width variation of Fe {\footnotesize II} (5991.4~{\AA}).
The dotted line shows the regression line with a slope of $(-5.0\pm1.2)\times10^{-5}$.
{\it Middle}: Equivalent width variation of Si {\footnotesize II} (6347.1~{\AA}).
The blue long-dashed line shows the regression line calculated with plots with MJD $\leq57500$, with a slope of $(-4.4\pm0.5)\times10^{-4}$.
The orange short-dashed line is the regression line of plots with MJD $>57500$, with a slope of $(-2.7\pm3.2)\times10^{-5}$.
{\it Bottom}: The variation of equivalent width ratio of Fe {\footnotesize II} (5991.4~{\AA}) and Si {\footnotesize II} (6347.1~{\AA}).
The blue long-dashed line is the regression line of plots with MJD $\leq57500$, with a slope of $(1.1\pm0.3)\times10^{-3}$.
The orange short-dashed line shows the regression line of plots with MJD $>57500$, with a slope of $(0.1\pm3.3)\times10^{-4}$. \label{fig:ew_ewr}}
\end{figure}

The equivalent width (EW) variations of the absorption lines reflect the physical parameters of V960 Mon spectrum.
The EWs of absorption lines were estimated with the ``e" command of the splot task in IRAF, by summing up the flux of an absorption line between both endpoints (from $w1$ to $w2$).
The error of each EW was calculated with the equation of \citet{Cayrel1988}, $\delta EW = 1.6(w\delta x)^{1/2}\epsilon$,
where $w$, $\delta x$, and $\epsilon$ are the width of the line, pixel size, and reciprocal of S/N, respectively.
The width of the line was estimated as $w_2 - w_1$.
Some of the absorption lines were not able to estimate the EW since they were located at the edge of the echelle aperture.
The lines which were contaminated by cosmicrays were also excluded from EW measurements.
The EWs of most of the lines were nearly consistent throughout the observation period (Figure~\ref{fig:ew}).
The measured EWs of Fe~{\footnotesize I} (6393.6~{\AA}, 6411.6~{\AA}) and Ca~{\footnotesize I} (6439.1~{\AA}) were comparable with the result of \citet{Takagi2018}.
Since the ratio of the broad component and the narrow component changed, EWs did not show a significant variation in these lines while the peak depth increased.
Meanwhile, EWs of relatively strong lines such as Ba~{\footnotesize II} (5853.7~{\AA}) and Ca~{\footnotesize I} (6122.2~{\AA}) showed a variation, especially between 2015 and 2017.
In addition, the EWs of Fe ~{\footnotesize II} (5991.4 {\AA}) and Si~{\footnotesize II} (6347.1 {\AA}) decreased, and the decreasing trend was not
consistent between these two lines (top and middle panels of Figure \ref{fig:ew_ewr}).
The slope of the regression line calculated with the EW of Fe~{\footnotesize II} (5991.4 {\AA}) was $(-5.0\pm1.2)\times10^{-5}$, which indicates that it gradually decreased from 2014 to 2020 February.
On the other hand, the slope of the regression line of Si~{\footnotesize II} (6347.1 {\AA}) was $(-4.4\pm0.5)\times10^{-4}$ from 2014 December to 2016 February,
indicating that the EW of Si~{\footnotesize II} decreased rapidly compared to Fe~{\footnotesize II} (5991.4 {\AA}). Then the EW of Si~{\footnotesize II} became constant after 2016 February.
The residual of the plots may imply the short-term variability of the EWs, similar to the periodic variation observed in FU Orionis \citep{Powell2012}.
The discrepancy in decline rates (Figure \ref{fig:ew_ewr} bottom) indicates that these decreases of EWs were not caused by the "filling-in" effect due to the variation of the continuum excess, so-called veiling.

\vspace{1cm}

\subsection{Cause of the variations in absorption lines} \label{subsec:absvar}

As described in sections \ref{subsec:absprof} and \ref{subsec:absEW}, the line profile, width, and EW were not uniform along the lines.
Based on the steady-disk model \citep{Hartmann1985}, a FUor spectrum is mainly composed of radiation from the accretion disk.
The absorption profiles seen in FUors such as boxy shape or a double peak correspond to the radiation from a rotating disk.
The characteristics of the V960 Mon spectrum observed at the beginning of the observation period were similar to those of typical FUors.
However, the absorption peak at around 0~km~${\mathrm s}^{-1}$ seen in the latter-phase spectrum is an unusual feature for FUors.

Because the brightness of V960 Mon decreased during the observation period, it can be considered that the fraction of the central-star spectrum gradually increased in the observed spectrum.
In 2014 December, V960 Mon was 2.9 mag ($2.5^{2.9}=14.3$ times) brighter in the $R$-band compared to the quiescent phase.
In this phase, the fraction of the central-star spectrum in the observed spectrum was 0.07 (1/14.3), and thus the disk spectrum was dominant.
In 2020 February, the brightness increment of V960 Mon was 1.4 mag in $R$-band ($2.5^{1.4}=3.6$ times).
Therefore, because of the fractional increase of the central star in the observed spectrum, the absorption peak at 0~km~${\mathrm s}^{-1}$ can be considered as the line of the central star,
which was unveiled as the disk spectrum faded.

\subsubsection{Line profile variations}\label{subsubsec:lineprof}

\begin{figure*}[t!]
 \begin{center}
   \includegraphics[width=16cm]{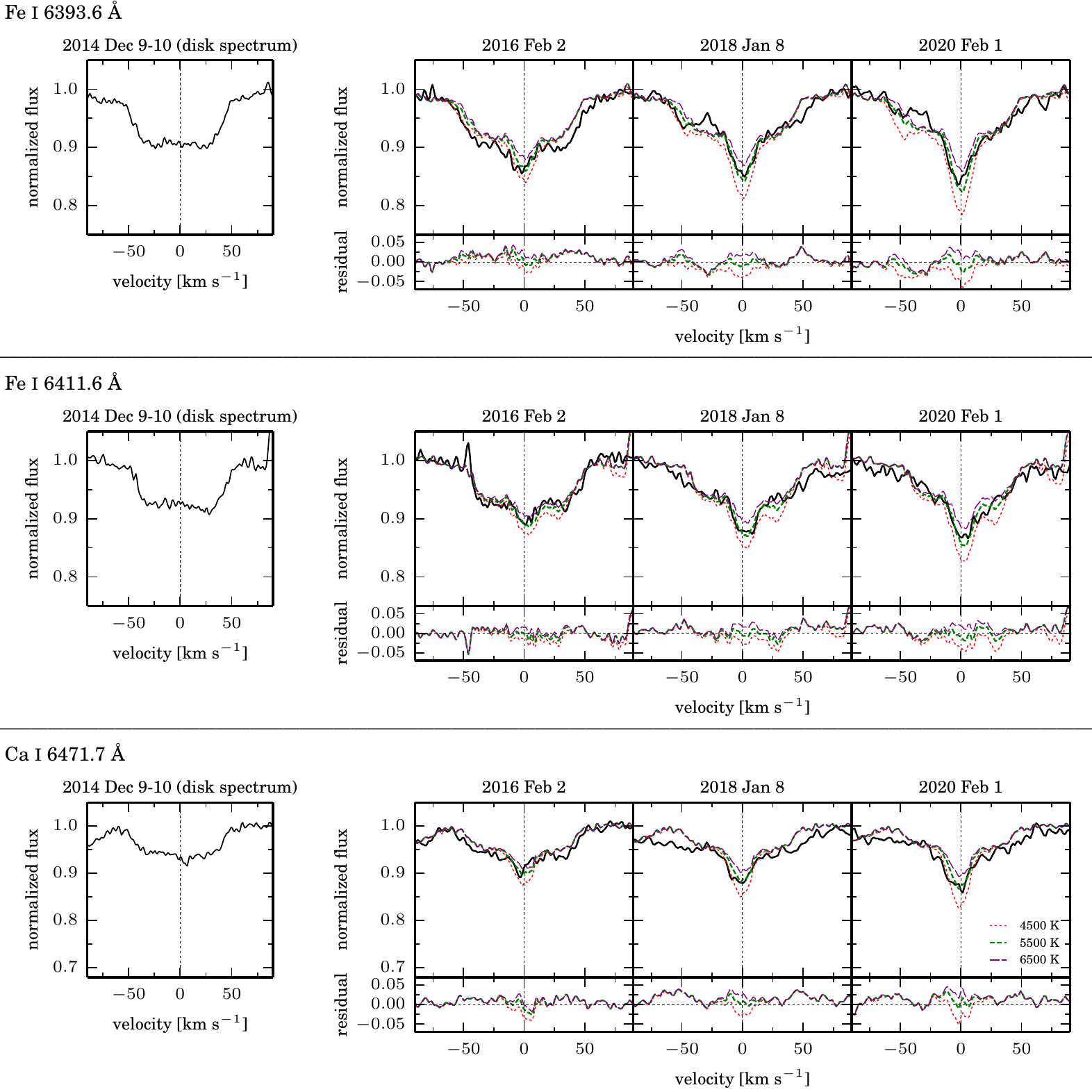}
 \end{center}
\caption{The result of model spectrum fitting. Left panels show the spectrum observed on 2014 December 9--10, which is assumed as the disk spectrum of V960 Mon in the model spectrum.
         Right panels show the comparison of the observed spectra (solid lines) and the model spectrum. The model spectrum is the sum of the disk spectrum (left panels) and the synthetic
	 spectrum reproducing the central-star.
	 Dotted line, short-dashed line, and long-dashed line represent model spectra with central-star $T_\mathrm{eff}$ of 4500 K, 5500K, and 6500 K, respectively.  \label{fig:compmodel}}
\end{figure*}

The spectrum of V960 Mon can be considered as a sum of central-star and disk-atmospheric spectra.
The physical parameter of the central star can be estimated by extracting the central-star component from the observed spectrum taken in the latter phase.
To estimate the $T_\mathrm{eff}$ and the log {\it g} of the central star, we established a simple model spectrum to reproduce the spectrum of 2016 February, 2018 January, and 2020 February.
The observed spectrum $I$ can be expressed as
\begin{equation}
  I=I_\mathrm{star}+kI_\mathrm{disk},
\end{equation}
where $I_\mathrm{star}$ is the central-star spectrum, $I_\mathrm{disk}$ is the disk-atmospheric
spectrum, and $k$ is the coefficient.
For $I_\mathrm{star}$, we used the software SPTOOL developed by Yoichi Takeda, which is based on Kurucz’s ATLAS9/WIDTH9 atmospheric model \citep{Kurucz1993}.
The synthetic spectra of the central star with $T_\mathrm{eff}$ of 4500, 5000, 5500, 6000, 6500, and 7000~K were created.
The log {\it g}, rotational velocity ($v$ sin $i$), and the microturbulence were fixed to typical values for a pre-main sequence star, which are 3.7, 10.0~km~${\mathrm s}^{-1}$, and 1.6~km~${\mathrm s}^{-1}$ \citep{Padgett1996}, respectively.
The metal abundance was set to the solar value.
For $I_\mathrm{disk}$, we used the observed spectrum of December 2014 \citep{Hillenbrand2014} for simplicity.
The coefficient $k$ was calculated based on the brightness.
The brightness increments of V960 Mon in 2016 February, 2018 January, and 2020 February were 2.1 mag, 1.7 mag, and 1.4 mag, respectively, in the $R-$band (see section \ref{sec:bri}).
Based on the brightness, the adopted $k$ values were 5.8 for February 2016, 3.7 for January 2018, and 2.6 for February 2020.

\renewcommand{\arraystretch}{0.85}
\begin{table}[t]
\caption{The sum of the squared residual (SSR) of the model spectrum and observed spectrum for each absorption line.} \label{tab:ssr}
\begin{center}
\footnotesize
\begin{tabular}{lrrrrrr}
  \hline \hline
  \multicolumn{1}{c}{}	& \multicolumn{6}{c}{SSR} \\
  \multicolumn{1}{c}{}	& \multicolumn{1}{c}{4500 K} & \multicolumn{1}{c}{5000 K} & \multicolumn{1}{c}{5500 K} & \multicolumn{1}{c}{6000 K} & \multicolumn{1}{c}{6500 K} & \multicolumn{1}{c}{7000 K} \\ \hline
  \multicolumn{7}{l}{2016 Feb 2} \\ \hline
  Ba~{\footnotesize II} 5853.7 {\AA}	& 0.095	& 0.107	& 0.119	& 0.127	& 0.133	& 0.140 \\
  Fe~{\footnotesize I} 5934.7 {\AA}	& 0.015	& 0.015	& 0.016	& 0.019	& 0.021	& 0.024 \\
  Fe~{\footnotesize I} 6003.0 {\AA}	& 0.021	& 0.017	& 0.016	& 0.016	& 0.017	& 0.018 \\
  Ca~{\footnotesize I} 6122.2 {\AA}	& 0.074	& 0.097	& 0.119	& 0.136	& 0.147	& 0.155 \\
  Fe~{\footnotesize I} 6265.1 {\AA}	& 0.098	& 0.085	& 0.079	& 0.077	& 0.078	& 0.080 \\
  Fe~{\footnotesize I} 6393.6 {\AA}	& 0.026	& 0.028	& 0.035	& 0.043	& 0.050	& 0.057 \\
  Fe~{\footnotesize I} 6411.6 {\AA}	& 0.031	& 0.024	& 0.019	& 0.019	& 0.020	& 0.022 \\
  Ca~{\footnotesize I} 6439.1 {\AA}	& 0.172	& 0.162	& 0.165	& 0.170	& 0.175	& 0.179 \\
  Ca~{\footnotesize I} 6471.7 {\AA}	& 0.027	& 0.022	& 0.020	& 0.021	& 0.022	& 0.023 \\
  Fe~{\footnotesize I} 6546.2 {\AA}	& -	& -	& -	& -	& -	& - \\
  Ni~{\footnotesize I} 7555.6 {\AA}	& 0.077	& 0.077	& 0.080	& 0.083	& 0.088	& 0.093 \\
  Fe~{\footnotesize I} 7568.9 {\AA}	& 0.039	& 0.039	& 0.036	& 0.033	& 0.031	& 0.031 \\
  Ni~{\footnotesize I} 7727.6 {\AA}	& 0.023	& 0.024	& 0.021	& 0.019	& 0.017	& 0.017 \\
  Fe~{\footnotesize I} 7937.1 {\AA}	& 0.032	& 0.032	& 0.029	& 0.028	& 0.028	& 0.029 \\
  $\Sigma$SSR \rule[-1.5mm]{0mm}{5mm}	& 0.730	& 0.729	& 0.754	& 0.791	& 0.827 & 0.868 \\ \hline
  \multicolumn{7}{l}{2018 Jan 8} \\ \hline
  Ba~{\footnotesize II} 5853.7 {\AA}	& 0.022	& 0.022	& 0.025	& 0.028	& 0.031	& 0.035 \\
  Fe~{\footnotesize I} 5934.7 {\AA}	& 0.015	& 0.014	& 0.016	& 0.019	& 0.024	& 0.028 \\
  Fe~{\footnotesize I} 6003.0 {\AA}	& 0.039	& 0.021	& 0.014	& 0.011	& 0.010	& 0.010 \\
  Ca~{\footnotesize I} 6122.2 {\AA}	& 0.042	& 0.027	& 0.044	& 0.061	& 0.074	& 0.084 \\
  Fe~{\footnotesize I} 6265.1 {\AA}	& 0.035	& 0.023	& 0.017	& 0.016	& 0.018	& 0.022 \\
  Fe~{\footnotesize I} 6393.6 {\AA}	& 0.033	& 0.019	& 0.015	& 0.018	& 0.023	& 0.028 \\
  Fe~{\footnotesize I} 6411.6 {\AA}	& 0.031	& 0.024	& 0.018	& 0.018	& 0.022	& 0.026 \\
  Ca~{\footnotesize I} 6439.1 {\AA}	& 0.061	& 0.023	& 0.016	& 0.018	& 0.021	& 0.024 \\
  Ca~{\footnotesize I} 6471.7 {\AA}	& 0.035	& 0.034	& 0.036	& 0.040	& 0.044	& 0.049 \\
  Fe~{\footnotesize I} 6546.2 {\AA}	& 0.054	& 0.035	& 0.018	& 0.013	& 0.015	& 0.023 \\
  Ni~{\footnotesize I} 7555.6 {\AA}	& 0.026	& 0.026	& 0.022	& 0.018	& 0.016	& 0.015 \\
  Fe~{\footnotesize I} 7568.9 {\AA}	& 0.019	& 0.019	& 0.016	& 0.015	& 0.014	& 0.015 \\
  Ni~{\footnotesize I} 7727.6 {\AA}	& 0.039	& 0.040	& 0.034	& 0.028	& 0.024	& 0.021 \\
  Fe~{\footnotesize I} 7937.1 {\AA}	& 0.052	& 0.044	& 0.031	& 0.022	& 0.018	& 0.016 \\
  $\Sigma$SSR \rule[-1.5mm]{0mm}{5mm}	& 0.503	& 0.371	& 0.322	& 0.325	& 0.354 & 0.396 \\ \hline
  \multicolumn{7}{l}{2020 Feb 1} \\ \hline
  Ba~{\footnotesize II} 5853.7 {\AA}	& 0.014	& 0.012	& 0.015	& 0.018	& 0.023	& 0.030 \\
  Fe~{\footnotesize I} 5934.7 {\AA}	& 0.018	& 0.016	& 0.015	& 0.018	& 0.022	& 0.028 \\
  Fe~{\footnotesize I} 6003.0 {\AA}	& 0.041	& 0.020	& 0.013	& 0.011	& 0.012	& 0.015 \\
  Ca~{\footnotesize I} 6122.2 {\AA}	& 0.064	& 0.018	& 0.037	& 0.060	& 0.078	& 0.093 \\
  Fe~{\footnotesize I} 6265.1 {\AA}	& 0.056	& 0.033	& 0.022	& 0.017	& 0.018	& 0.023 \\
  Fe~{\footnotesize I} 6393.6 {\AA}	& 0.056	& 0.032	& 0.018	& 0.018	& 0.025	& 0.033 \\
  Fe~{\footnotesize I} 6411.6 {\AA}	& 0.036	& 0.023	& 0.014	& 0.015	& 0.020	& 0.026 \\
  Ca~{\footnotesize I} 6439.1 {\AA}	& 0.055	& 0.032	& 0.041	& 0.055	& 0.068	& 0.079 \\
  Ca~{\footnotesize I} 6471.7 {\AA}	& 0.029	& 0.024	& 0.027	& 0.033	& 0.039	& 0.047 \\
  Fe~{\footnotesize I} 6546.2 {\AA}	& 0.119	& 0.082	& 0.043	& 0.027	& 0.030	& 0.046 \\
  Ni~{\footnotesize I} 7555.6 {\AA}	& 0.042	& 0.041	& 0.036	& 0.031	& 0.028	& 0.028 \\
  Fe~{\footnotesize I} 7568.9 {\AA}	& 0.043	& 0.043	& 0.036	& 0.032	& 0.030	& 0.029 \\
  Ni~{\footnotesize I} 7727.6 {\AA}	& 0.047	& 0.049	& 0.040	& 0.031	& 0.024	& 0.021 \\
  Fe~{\footnotesize I} 7937.1 {\AA}	& 0.048	& 0.041	& 0.027	& 0.019	& 0.016	& 0.016 \\
  $\Sigma$SSR  \rule[-1.5mm]{0mm}{5mm}	& 0.668	& 0.466	& 0.384	& 0.385	& 0.433 & 0.514 \\ \hline
 \hline
 \end{tabular}
 \end{center}
\end{table}
\renewcommand{\arraystretch}{1}

\begin{figure*}[t!]
 \begin{center}
   \includegraphics[width=16cm]{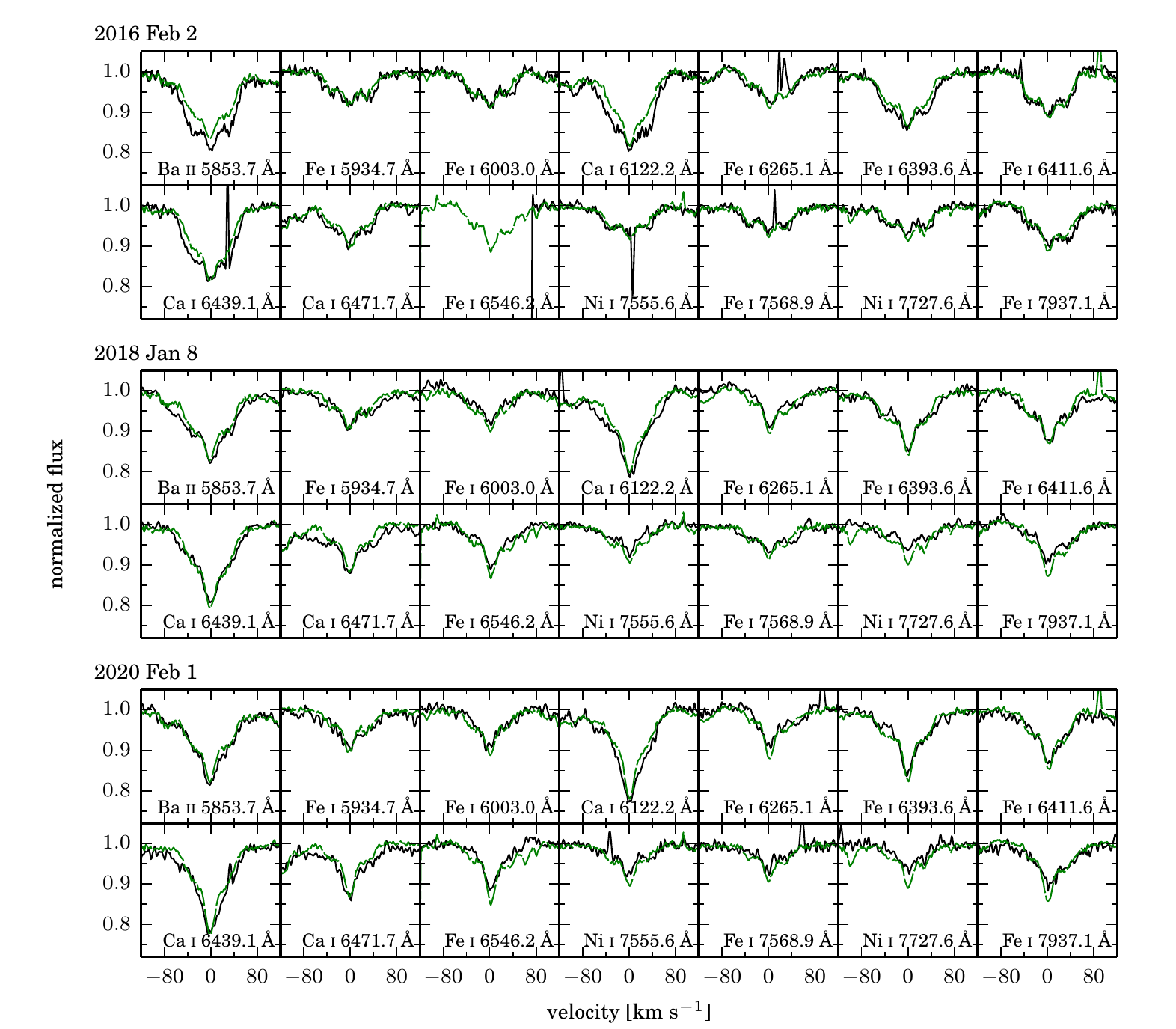}
 \end{center}
\caption{Comparison of the line profiles of observed spectra and the model spectrum with the central star of 5500 K (green dashed line). \label{fig:lcwwm}}
\end{figure*}

We compared the profile of absorption lines with negligibly small or no blending with other lines (Table \ref{tab:linedata}) between the model spectra and the observed spectra (Figure \ref{fig:compmodel}).
The sum of the squared residual (SSR) was calculated in each absorption line (Table \ref{tab:ssr}).
The sum of SSR for all lines ($\Sigma$SSR) became minimum when the effective temperature of the central star was set to 5000 -- 6000~K.
We also created a model spectrum that takes into account the brightness variation of V960 Mon which leads to the change in the brightness ratio of the disk and central star.
The brightness variability of the quiescent phase was assumed as $\pm 0.3$ mag based on the pre-outburst brightness \citep{Kospal2015}.
The model spectrum including the brightness uncertainty was compared with the observed spectrum on 2018 February.
Even including this brightness uncertainty, the $\Sigma$SSR became minimum when the $T_\mathrm{eff}$ was 5500 -- 6000~K.

The $T_\mathrm{eff}$ estimated from the observed spectrum conflicts with that of pre-outburst SED \citep[4000~K, ][]{Kospal2015}, which was calculated on the assumption that there are no extinction and a distance is 450~pc.
On the other hand, the distance estimated by Gaia is $1.6\pm0.2$~kpc \citep{Gaia, GaiaDR2}, which implies that the absolute magnitude of V960 Mon is brighter, and therefore the $T_\mathrm{eff}$ can be higher.
However, the V960 Mon should be reddened to explain the pre-outburst SED with $T_\mathrm{eff}$ higher than 4000~K.

We investigated whether the progenitor photosphere of 5500~K is compatible with the pre-outburst brightness and the distance of 1.6~kpc.
To estimate the extinction, we compared the pre-outburst color with the typical color of a star with a $T_\mathrm{eff}$ of 5500~K.
The color of $I-J$ was used since the $I$-band brightness is likely to be most less effected by the veiling \citep{Cieza2005}.
The averaged $I-$ and $J$-band magnitudes of \citet{Kospal2015} were used for calculations, which were $I=12.17\pm0.35$ mag and $J=11.03\pm0.02$ mag,
and then the pre-outburst $I-J$ is calculated as $1.14\pm0.35$ mag.
By comparing this color with the typical color of 5500~K photosphere \citep[0.47 mag, ][]{KH95} and the reddening law \citep{Mathis1990}, the extinction in $I$-band was estimated as $1.26\pm0.66$ mag.
The pre-outburst SED can be explained well with a central star of $T_\mathrm{eff}=5500$~K and the extinction of 1.26 mag with a near-infrared excess typically seen in young stars.
The absolute magnitude of V960 Mon in $I$-band can be calculated as $-0.11\pm0.45$ mag, taking into account the distance estimated by Gaia and assuming the extinction in $I$-band as 1.26 mag.
This magnitude is comparable to the absolute magnitude of a pre-main sequence star with $T_\mathrm{eff}$ of 5500~K and the age of $\sim$0.7 Myr \citep{Siess2000}.
Therefore, the photosphere with the $T_\mathrm{eff}$ of 5500~K is comparable with the pre-outburst brightness and the distance of Gaia.
However, when the extinction is an amount of lower or upper limit, the SED cannot be explained with a 5500~K central star.
To discuss the central-star nature, further observations and discussions of the post-outburst phase are necessary.

In addition, there were several discrepancies between the observed spectrum and the model spectrum.
The absorption lines in the wavelength of $>$7000 {\AA} showed a slight mismatch between the observation and the model, especially in the spectrum of 2018 January 8.
The broad component of the model spectrum is deeper than that compared to the observed spectrum.
The coefficient $k$ used in the model calculation is determined from the $R-$band magnitude.
Because the decrease rate of the brightness $I-$band was lower compared to the $R-$band (Figure \ref{fig:lightcurve}), the fraction of the disk spectrum in the observed spectrum may be larger in long wavelength.
The broad components of relatively strong lines such as Ba~{\footnotesize II} (5853.7 {\AA}) and Ca~{\footnotesize II} (6122.2 {\AA}) lines also show a mismatch.
Especially, these lines in 2016 February 2 were deeper than those of the model spectrum.
Because the $I_\mathrm{disk}$ used in the model is the observed spectrum of 2014 December, our model spectrum does not take into account the variation of the disk spectrum during its fading.
The differences between the model spectrum and the observed spectrum may result from the change in the disk nature.

\subsubsection{Fe~{\footnotesize II} and Si~{\footnotesize II} lines and the disk spectrum}\label{subsubsec:ionline}

\begin{figure}[t!]
 \begin{center}
   \includegraphics[width=8cm]{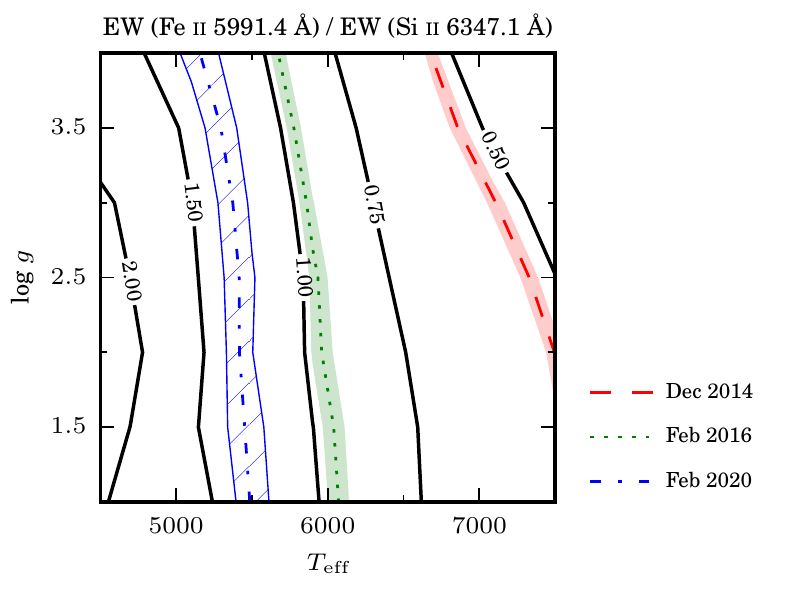}
 \end{center}
\caption{Contour plot showing the relation between $T_\mathrm{eff}$, log {\it g}, and the EW ratio of Fe~{\footnotesize II} (5991.4 {\AA}) and Si~{\footnotesize II} (6347.1 {\AA})
         to estimate the variation of physical parameters of disk atmosphere.
         Red dashed line, green dotted line, and the blue dash-dotted line show the EW ratio of 2014 December, 2016 February, and February 2020, respectively.
         The shaded areas show the error of each EW ratio. \label{fig:ewr_model}}
\end{figure}

\begin{figure}[h!]
 \begin{center}
   \includegraphics[width=8cm]{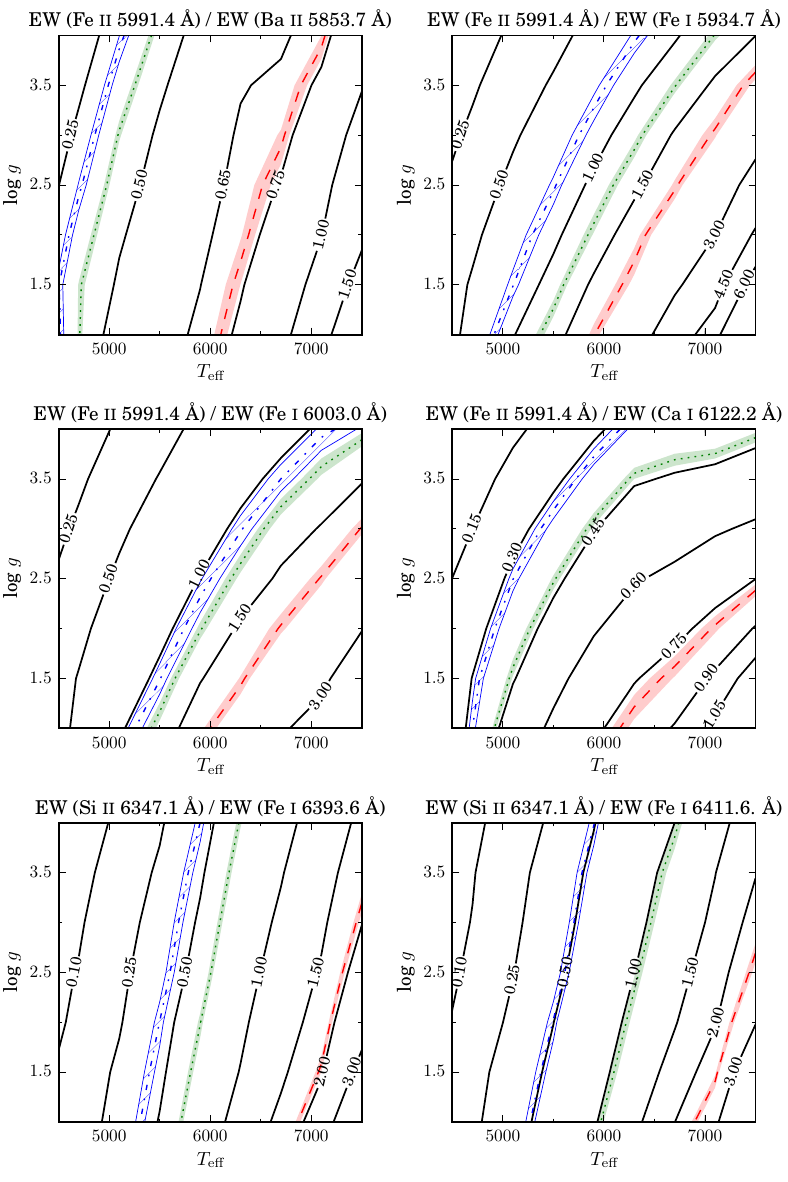}
 \end{center}
\caption{Contour plot showing the relation between $T_\mathrm{eff}$, log {\it g}, and the EW ratios, to estimate the variation of physical parameters of disk atomsphere.
         Red dashed line, green dotted line, and the blue dash-dotted line show the EW ratios of 2014 December, 2016 February, and February 2020, respectively,
         estimated from the spectrum whose central-star component was subtracted.
         The shaded areas show the error of each EW ratio. \label{fig:ewr_model_forsubsp}}
\end{figure}

While most of the absorption lines show a change in line profile, Fe~{\footnotesize II} (5991.4 {\AA}) and Si~{\footnotesize II} (6347.1 {\AA}) did not show profile variations during the observation period.
This indicates that the central-star spectrum does not have strong Fe~{\footnotesize II} and Si~{\footnotesize II} lines.
In fact, these lines are weak in the star with a $T_\mathrm{eff}$ of 5500~K.
Therefore, the EW variations of these two lines reflect the change in the physical parameters of the disk atmosphere.

Because the observed spectrum of V960 Mon can be assumed to be a sum of the disk spectrum and central-star spectrum, the EW of an absorption line includes the information of both disk and the central-star atmosphere.
The EWs of the Fe~{\footnotesize II} and Si~{\footnotesize II} lines can be considered that they are veiled by the stellar continuum of the central star.
For this reason, it is difficult to derive the disk property from the EW itself.

The EW ratio of nearby absorption lines is a valid tool for estimating the physical parameters of the spectrum with continuum veiling \citep{Takagi2010, Takagi2011}.
The ``filling in" effect of absorption lines caused by the veiling can be removed by using the EW ratio of nearby absorption lines by assuming that the veiling is nearly uniform among absorption lines in a limited wavelength range.
We used the EW ratio of Fe~{\footnotesize II} and Si~{\footnotesize II} lines to estimate the variations of $T_\mathrm{eff}$ and log {\it g} in the disk atmospheric spectrum.
We first investigated the relation between the $T_\mathrm{eff}$, log {\it g}, and the EW ratio by using the synthetic spectrum.
The synthetic spectra were created by changing the $T_\mathrm{eff}$ from 4500 K to 7500 K, and log {\it g} from 1.0 to 4.0, by using the software SPTOOL.
The metal abundance was set to the solar value.
Microturbulence was set to a value based on the empirical relations \citep{Gray2001, Takagi2018}.

A comparison of the EW ratio of the observed spectrum (Figure \ref{fig:ew_ewr} bottom) and the derived EW--$T_\mathrm{eff}$--log {\it g} relationship showed that
the physical parameters of the disk spectrum changed during the observation period (Figure \ref{fig:ewr_model}).
The EW ratio of Fe~{\footnotesize II} and Si~{\footnotesize II} increased by a factor of two from 2014 December to 2016 February.
This EW ratio variation indicates that $T_\mathrm{eff}$ decreased by $\sim$1000~K during this period.
Meanwhile, it is difficult to estimate the variation in log~{\it g}.
The variation from 2016 February to 2020 February was uncertain since the EW ratio of Fe~{\footnotesize II} and Si~{\footnotesize II} was nearly stable during this period (Figure \ref{fig:ew_ewr}).

We also used other absorption lines such as Fe~{\footnotesize I} and Ca~{\footnotesize I} to discuss the variations in the disk atmosphere.
The spectrum of the disk atmosphere can be extracted from the observation spectrum by subtracting the central-star component.
The disk components of 2016 February and 2020 February were obtained by subtracting the synthetic spectrum which reproduces the central star.
The $T_\mathrm{eff}$ of this synthetic spectrum was set to 5500~K, based on the model fitting (section \ref{subsubsec:lineprof}).
Other parameters such as log~{\it g}, $v$~sin~$i$, metallicity, and microturbulence velocity were the same as described in section \ref{subsubsec:lineprof}.
The fractions of the central-star spectrum and the disk spectrum were calculated based on the brightness of V960 Mon (section \ref{sec:bri}).
We used relatively strong absorption lines for the EW ratio, which were
Ba~{\footnotesize II} (5853.7~{\AA}),
Fe~{\footnotesize I} (5934.7~{\AA}),
Fe~{\footnotesize II} (5991.4~{\AA}),
Fe~{\footnotesize I} (6003.0~{\AA}),
Ca~{\footnotesize I} (6122.2~{\AA}),
Si~{\footnotesize II} (6347.1~{\AA}),
Fe~{\footnotesize I} (6393.6~{\AA}), and
Fe~{\footnotesize I} (6411.6~{\AA}).
The EW ratios were calculated by combining Fe~{\footnotesize II} (5991.4~{\AA}) and Si~{\footnotesize II} (6347.1~{\AA}) with other lines.
Absorption lines within a limited wavelength range ($\sim$200 {\AA}) were used for EW ratios.

All six pairs of EW ratios (Figure \ref{fig:ewr_model_forsubsp}) showed that the physical parameters of the disk atmosphere changed during the dimming event.
By taking into account the variation of the EW ratio of Fe~{\footnotesize II} and Si~{\footnotesize II}, it can be assumed that the changes of these EW ratios
were caused by the decrease in $T_\mathrm{eff}$.
This indicates that the accretion rate decreased \citep{Kenyon1988,Kenyon1991}, which is consistent with the decline in the brightness.
The change in the V960 Mon color is also consistent with this cooling.
According to the photometric observations (Figure \ref{fig:lightcurve}), $B-V$, $V-R$, and $R-I$ on 2014 December 19 were 1.0 mag, 0.8 mag, and 0.7 mag, respectively.
These colors then changed to 1.5 mag, 0.9 mag, and 1.0 mag on 2020 February 1.

The change in the physical parameter of the disk can be considered as one of the causes of the discrepancy in the model spectrum and the observed spectrum,
especially seen in the Ba~{\footnotesize II} (5853.7 {\AA}) and Ca~{\footnotesize I} (6122.2 {\AA}) lines (Figure \ref{fig:lcwwm}).
Meanwhile, even using the disk component spectrum, the actual $T_\mathrm{eff}$ and log~{\it g} values of the disk atmosphere are difficult to estimate.
One of the causes of this difficulty may be the uncertainty of the physical parameter estimations of the central star.
Another reason is the variations in the local disk atmosphere cannot be revealed with the EW ratio investigations.
The disk spectrum is considered to be composed of the disk atmosphere which extends from the hot inner region to the cold outer part of the disk \citep{Hartmann1996}.
The observed disk spectrum is the sum of the spectra arising from different regions with various $T_\mathrm{eff}$.
Therefore, the region where an absorption line arises dominantly may depend on each line.
In fact, the line widths of Fe~{\footnotesize II} and Si~{\footnotesize II} has decreased as V960 Mon faded, and the decreasing rate was not equal between these two lines (Figure \ref{fig:fw15d}).
Investigations of EW ratios can reveal the averaged parameters of the whole disk atmosphere, which allows a qualitative study of the physical parameter variations.

\subsection{H$\alpha$ and Ca~{\footnotesize II} lines} \label{subsec:emi}

\begin{figure*}[t!]
 \begin{center}
   \includegraphics[width=18cm]{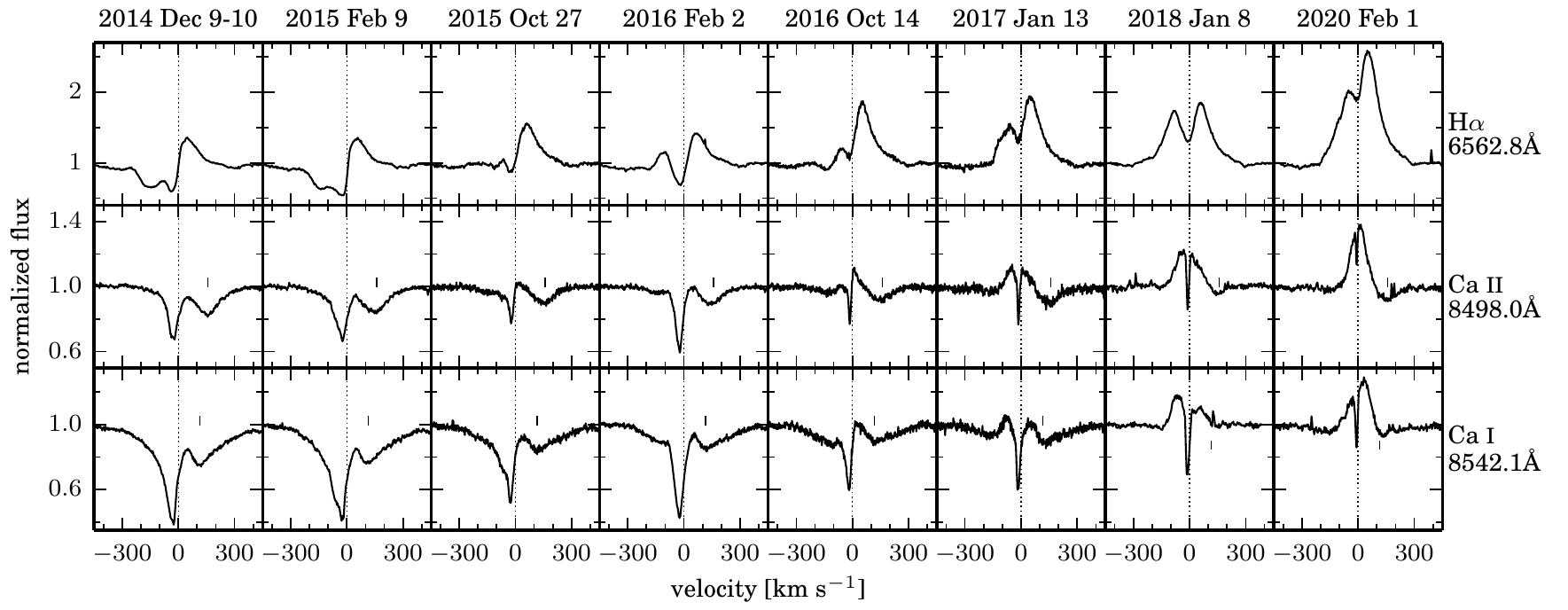}
 \end{center}
\caption{Variations of the H$\alpha$ and Ca~{\footnotesize II} lines. Short vertical lines in middle and bottom rows show the wavelength of Paschen absorption series. \label{fig:lcs}}
\end{figure*}

\begin{figure}[t!]
 \begin{center}
   \includegraphics[width=8cm]{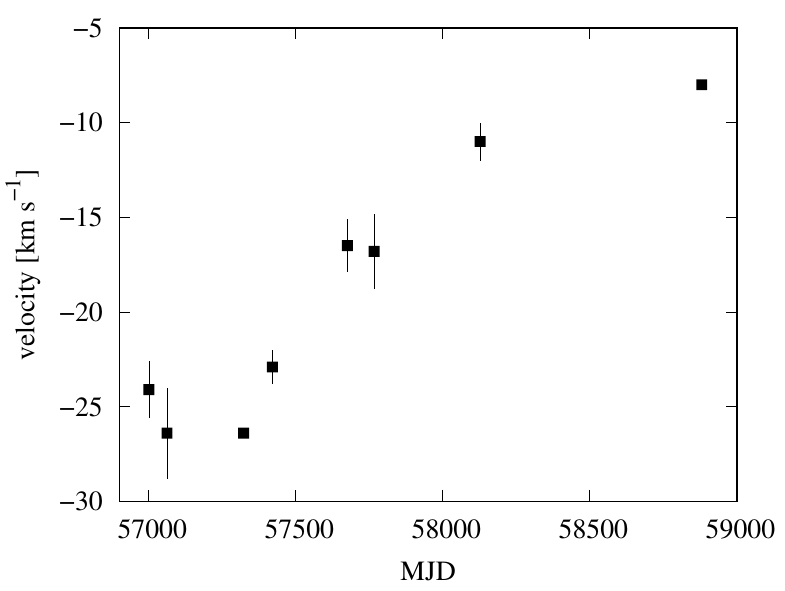}
 \end{center}
\caption{The averaged peak velocity of Ca II absorption lines. \label{fig:CaIIshift}}
\end{figure}

Strong lines such as H$\alpha$ and Ca~{\footnotesize II} lines also showed significant variations during the observation period.
Both H$\alpha$ and Ca~{\footnotesize II} lines had absorption components, especially in the early phase of the observation.
It can be considered that the absorption component at the line center arises from the disk atmosphere.
The blueshifted absorption component is evidence of an active outflow.
As V960 Mon faded, emission features became dominant instead of the absorption features.
The increase of the emission features suggests that the typical emission lines of a pre-main sequence star reappear, because the fraction of the disk atmosphere has decreased.

Ca~{\footnotesize II} lines (8498.0~{\AA}, 8542.1~{\AA}) were blended with the Paschen lines (8502.5 {\AA}, 8545.4 {\AA}).
Other Paschen lines (8438.0~{\AA}, 8467.3~{\AA}) were also detected, which were blended with the nearby Fe~{\footnotesize I} lines.
The Paschen absorption lines are seen in other FU Ori type stars \citep{Connelley2018}.
It was difficult to investigate the variation of the Paschen lines due to the blend, but they became shallow in the latter phase of the observation.
These lines are seen in stars with high $T_\mathrm{eff}$
($<$ F type star) and small gravity such as giants. Therefore, the depth decrease of these lines is consistent with the change of EW ratios
(section \ref{subsubsec:ionline}), which indicates the $T_\mathrm{eff}$ decrease of the disk atmosphere. In addition, these lines are
also veiled by the central-star spectrum mainly in the early phase of the observation period.

The absorption peak of Ca~{\footnotesize II} lines showed a shift. The averaged peak velocity was $\sim-25$ km~${\mathrm s}^{-1}$ from
2014 -- 2016, and then it decreased to $-8$ km~${\mathrm s}^{-1}$ in 2020 February. This result indicates that the blueshifted absorption
component from the outflow became weak. The shift of the absorption peak is also consistent with the trend that the spectrum of the
central star became dominant in the latter phase.

\section{Summary} \label{sec:sum}

We conducted a study of spectroscopic variations of the FU Ori-type star V960 Mon.
We used high-resolution optical spectra from 2014 December to 2020 February, including both observed and archival data during the phase in which the brightness of V960 Mon decreased by 1.5 mag in the $R$-band.
The depths of the neutral and ionized absorption lines showed different variations; most of the lines became deep while the ionized lines such as Fe~{\footnotesize II} (5991.4 {\AA}) and Si~{\footnotesize II} (6347.1 {\AA}) became shallow.
In addition, the profile of the absorption lines such as Fe~{\footnotesize I} and Ca~{\footnotesize I} changed.
The absorption profile shown in the former phase was mainly composed of the disk atmosphere.
As V960 Mon faded, the component of the central star became dominant in the absorption line profile.
Our model spectrum showed that the observed spectrum can be described by a sum of the spectrum from the disk atmosphere and that from the central star, where the $T_\mathrm{eff}$ of the central star can be estimated as 5500~K.
The disk spectrum also indicated variation in the $T_\mathrm{eff}$.
The EW ratios of the absorption lines presumed that the averaged $T_\mathrm{eff}$ decreased by $~1000$~K.
The H$\alpha$ and Ca~{\footnotesize II} lines also showed variations due to the change in the fractions of two components, the disk atmosphere and the central star.

\medskip

This work was supported by JSPS KAKENHI Grant Numbers 19K03957 and 26870507. We acknowledge with thanks the variable star
observations from the AAVSO International Database contributed by observers worldwide and used in this research.
This research has made use of the Keck Observatory Archive (KOA), which is operated by the W. M. Keck Observatory and the NASA Exoplanet
Science Institute (NExScI), under contract with the National Aeronautics and Space Administration.
This work has made use of data from the European Space Agency (ESA) mission {\it Gaia} (\url{https://www.cosmos.esa.int/gaia}),
processed by the {\it Gaia} Data Processing and Analysis Consortium (DPAC, \url{https://www.cosmos.esa.int/web/gaia/dpac/consortium}).
Funding for the DPAC has been provided by national institutions, in particular the institutions participating in the {\it Gaia} Multilateral Agreement.
%This publication makes use of data products from the Two Micron All Sky Survey, which is a joint project of the University of Massachusetts
%and the Infrared Processing and Analysis Center/California Institute of Technology,
%funded by the National Aeronautics and Space Administration and the National Science Foundation.

\software{IRAF \citep{Tody1986,Tody1993}}.

\bibliographystyle{aasjournal}
\bibliography{reference}

\end{document}